%% file: main.tex
\documentclass[sigconf]{acmart}

\acmConference[VibeX 2026]{1st International Workshop on Vibe Coding and Vibe Researching}{9--12 June, 2026}{Glasgow, Scotland, United Kingdom}

\renewcommand\footnotetextcopyrightpermission[1]{}
\usepackage{booktabs}
\usepackage{tikz}
\usetikzlibrary{arrows.meta,positioning,shapes.geometric,fit,backgrounds}
\usepackage{orcidlink}

\title{Mise en Place for Agentic Coding: Deliberate Preparation as Context Engineering Methodology}

\author{Andrew Zigler\,\orcidlink{0009-0001-8073-5917}\,\href{https://orcid.org/0009-0001-8073-5917}{0009-0001-8073-5917}}
\email{andrew.zigler@linearb.io}
\affiliation{%
  \institution{LinearB}
  \city{Los Angeles}
  \country{USA}
}

\begin{document}

\begin{abstract}
The rapid adoption of AI coding agents has produced a dominant workflow pattern --- often called ``vibe coding'' --- that prioritizes speed of implementation over deliberate preparation. We argue that this approach creates a systematic alignment problem: agents that lack sufficient context produce code requiring extensive debugging and refactoring, consuming substantial development time. Drawing on the culinary concept of \emph{mise en place} (everything in its place; abbreviated MEP), we propose a three-phase preparation methodology for agentic coding: (1) contextual grounding, where domain expertise and tacit knowledge are externalized into structured documents; (2) collaborative specification, where human-agent dialogue produces detailed design artifacts; and (3) task decomposition, where specifications are converted into structured, dependency-aware task records. We report on the application of MEP during a competitive hackathon, where roughly two hours of preparation enabled a rapid parallel implementation of a full-stack educational platform by concurrent AI agents. We introduce the concept of \emph{context fluency} as an emerging developer skill --- the ability to create rich, structured context that agents can act on --- and connect it to established frameworks in backward design and tacit knowledge externalization. We conclude with a research agenda for empirically validating preparation-phase methodologies in AI-assisted software development.
\end{abstract}



\maketitle

\input{sections/introduction}
\input{sections/related-work}
\input{sections/methodology}
\input{sections/case-study}
\input{sections/context-fluency}
\input{sections/research-agenda}

\bibliography{references}

\end{document}

%% file: sections/introduction.tex
\section{Introduction}

AI coding agents are remarkably capable at code generation: controlled
studies of GitHub Copilot report productivity gains of 21--55\% across
programming tasks~\cite{peng2023copilot}, and the broader ecosystem can
now scaffold entire applications from natural-language descriptions.
Yet the dominant workflow pattern---what
Karpathy~\cite{karpathy2025vibe} termed ``vibe coding''---prioritizes
speed of implementation over deliberate preparation. The developer
describes an intent, the agent produces code, and misalignments are
resolved through iterative correction. This creates a systematic
alignment problem: agents operating without sufficient context produce
code that requires extensive rework. Veracode reports that 45\% of
AI-generated code contains security flaws~\cite{veracode2025ai}, and
practitioners consistently observe that debugging misaligned agent
output can consume substantial development
time~\cite{huntley2025ralph, yegge2025beads}. The bottleneck in agentic
coding is not code generation but \emph{alignment} between developer
intent and agent output---specification compliance, architectural
fidelity, and low corrective-commit ratios.

We argue that this alignment problem is fundamentally a \emph{preparation}
problem. Drawing on the culinary concept of \emph{mise en place} (MEP)---a
French term meaning ``everything in its place''---we propose a
preparation-first methodology for agentic coding. In professional kitchens,
thorough preparation enables fluid execution: every ingredient is
measured and every tool positioned so that once cooking begins, the
chef's hands never pause to search. The same principle applies to agent
orchestration. When domain expertise, design intent, and task boundaries
are externalized into structured artifacts \emph{before} agents begin
writing code, the resulting implementation is more aligned, more
coherent, and less costly to verify. MEP formalizes this preparation
into three steps: contextual grounding (tacit knowledge captured in
machine-legible documents), collaborative specification (human-agent
dialogue producing detailed design artifacts), and task decomposition
(specifications converted into dependency-aware work records). We
position MEP relative to spec-driven development, prompt engineering,
and iterative AI coding in Section~\ref{sec:related}.

This paper makes four contributions:
\begin{enumerate}
  \item A three-phase MEP methodology for agentic coding, grounded
    in backward design~\cite{wiggins1998understanding} and tacit knowledge
    externalization~\cite{polanyi1966tacit}.
  \item A case study applying this methodology in a competitive hackathon,
    with quantitative data on preparation artifacts and implementation
    outcomes.
  \item The concept of \emph{context fluency} as an emerging developer
    skill---the ability to create rich, structured context that AI agents can
    act on.
  \item A research agenda with five open questions for empirical validation
    of preparation-phase methodologies.
\end{enumerate}

%% file: sections/related-work.tex
\section{Background and Related Work}
\label{sec:related}

Early empirical studies of AI-assisted coding reported substantial productivity
gains: Peng et al.\ found that developers using GitHub Copilot completed tasks
21--55\% faster~\cite{peng2023copilot}. However, speed has not translated cleanly
into quality. Veracode's 2025 analysis found that 45\% of AI-generated code
contains security flaws~\cite{veracode2025ai}, while Mozannar et al.\ showed
that developers' mental models frequently diverge from actual AI behavior, leading
to inappropriate acceptance of incorrect suggestions~\cite{developer2023mental}.
Vasconcelos et al.\ demonstrated that trust in AI code generation depends on
context-aware explanations that current tools rarely
provide~\cite{facct2024trust}. Together, these findings point to a productivity paradox:
AI agents accelerate output while simultaneously increasing the surface area for
rework and defects. The bottleneck in agentic coding, we argue, is not code
generation but \emph{alignment}---ensuring that what agents build matches what
practitioners intend.

\subsection{Prompt and Context Engineering}

A substantial body of work studies how to elicit better behavior from
large language models through careful instruction. Foundational results
established that LLMs perform new tasks from in-context
examples~\cite{brown2020gpt3}, and that intermediate reasoning steps
improve task accuracy~\cite{wei2022chain}. Liu et al.'s
survey~\cite{liu2023prompt} catalogues a maturing taxonomy of prompting
techniques. More recently, the discourse has shifted from prompt
engineering to \emph{context engineering}: designing the complete
informational environment surrounding an LLM~\cite{anthropic2025context,
gartner2025context}. Alomar et al.\ extend this thread with
\emph{promptware engineering}~\cite{alomar2025promptware}, and Dakhel
et al.'s taxonomy identifies eleven distinct human-AI interaction
types~\cite{dakhel2025humanai}. MEP operates at a different scope from
prompt engineering: prompt engineering tunes individual model
invocations, whereas MEP structures the workflow-level artifacts
(specifications, decompositions, externalized tacit knowledge) that
precede and constrain those invocations. Both shape what the model
receives, but at different granularities.

\subsection{Preparation Methodologies}

Several preparation-oriented methodologies anticipate aspects of our
proposal. Horthy's Research-Plan-Implement (RPI)
methodology~\cite{horthy2025rpi} most directly approaches this work,
requiring practitioners to complete discrete research and planning
phases before any code is generated. MEP differs from RPI by
formalizing tacit knowledge externalization in Phase~1 and insisting on
backward-design sequencing where all three phases complete before
implementation begins. GitHub's Spec Kit~\cite{github2025speckit}
formalizes specification-driven development, where a structured
specification artifact governs agent behavior throughout
implementation. These contributions recognize that what happens
\emph{before} agent execution shapes outcomes, but none offers a
unified, theoretically grounded preparation methodology that
integrates tacit knowledge externalization, collaborative
specification, and dependency-aware decomposition into one phase-gated
sequence.

\subsection{Practitioner Tooling and Theoretical Grounding}

A parallel stream of practitioner innovation addresses preparation
through tooling. Karpathy coined ``vibe coding''~\cite{karpathy2025vibe}
for a minimal-preparation workflow scoped to throwaway projects, yet the
term came to characterize the dominant mode of AI-assisted development.
At the opposite end, Huntley's Ralph Loop~\cite{huntley2025ralph}
provides fine-grained iterative task delegation, and Yegge's Beads
framework~\cite{yegge2025beads} addresses agent memory through
structured task records, extended by Emanuel's Rust
reimplementation~\cite{emanuel2025beadsrust}. These tools solve specific
problems---iteration, memory, orchestration---but assume that
appropriate context already exists. MEP draws on two theoretical
frameworks. Wiggins and McTighe's
\emph{backward design}~\cite{wiggins1998understanding} holds that
effective instruction begins with desired outcomes, then works backward
through assessment criteria---we apply this principle by defining
agent outputs and success criteria before designing context. Polanyi's
tacit knowledge---``we can know more than we can
tell''~\cite{polanyi1966tacit, polanyi1958personal}---frames preparation
as a knowledge externalization problem; Nonaka and Takeuchi's SECI
model~\cite{nonaka1995knowledge} formalizes this tacit-to-explicit
conversion as a repeatable organizational process, grounding Phase~1.

\subsection{Positioning: What is New}
\label{sec:novelty}

The ingredients of MEP are individually well-established; our
contribution is integrative and scope-specific. Against
\emph{spec-driven development}~\cite{github2025speckit}, MEP adds
Phase~1 for externalizing tacit, value-laden knowledge that
specifications alone do not capture. Against \emph{prompt
engineering}~\cite{brown2020gpt3, wei2022chain, liu2023prompt}, MEP
operates at workflow rather than invocation scope. Against
\emph{iterative or vibe-coding workflows}~\cite{karpathy2025vibe,
huntley2025ralph}, MEP front-loads alignment work that iterative flows
pay incrementally as rework. Agentic workflows need this preparation
because agents lack the tacit context human collaborators carry and
cannot iterate cheaply without expensive re-generation.

%% file: sections/methodology.tex
\section{The Mise en Place Methodology}
\label{sec:methodology}

\begin{figure}[t]
  \centering
  \begin{tikzpicture}[
    >=Stealth,
    phase/.style={draw, rounded corners, fill=black!8, minimum width=2.1cm,
      minimum height=0.7cm, font=\footnotesize, align=center, text width=2cm},
    agent/.style={draw, rounded corners, fill=black!4, minimum width=1.2cm,
      minimum height=0.55cm, font=\scriptsize, align=center},
    phaselbl/.style={font=\scriptsize\bfseries, text=black!50},
    stagelbl/.style={font=\scriptsize\itshape, text=black!60},
    arw/.style={->, thick},
  ]
    \node[phaselbl] (l1) at (0, 0.55) {Phase 1};
    \node[phaselbl] (l2) at (2.8, 0.55) {Phase 2};
    \node[phaselbl] (l3) at (5.6, 0.55) {Phase 3};

    \node[phase] (p1) at (0,0) {Contextual\\Grounding};
    \node[phase] (p2) at (2.8,0) {Collaborative\\Specification};
    \node[phase] (p3) at (5.6,0) {Task\\Decomposition};

    \draw[arw] (p1) -- (p2);
    \draw[arw] (p2) -- (p3);

    \node[stagelbl] at (-1.5, 0) {\rotatebox{90}{Preparation}};

    \node[agent, below=1.1cm of p1, xshift=0.4cm] (a1) {Agent$_1$};
    \node[agent, right=0.25cm of a1] (a2) {Agent$_2$};
    \node[font=\footnotesize, right=0.15cm of a2] (dots) {$\cdots$};
    \node[agent, right=0.15cm of dots] (an) {Agent$_n$};

    \draw[arw] (p3.south) -- ++(0,-0.3) -| (a1.north);
    \draw[arw] (p3.south) -- ++(0,-0.3) -| (a2.north);
    \draw[arw] (p3.south) -- ++(0,-0.3) -| (an.north);

    \node[stagelbl] at (-1.5, -1.45) {\rotatebox{90}{Execution}};

    \node[phase, below=1.1cm of a2, xshift=0.35cm, minimum width=2.8cm] (ver) {Integration\\Verification};

    \draw[arw] (a1.south) -- ++(0,-0.3) -| ([xshift=-0.4cm]ver.north);
    \draw[arw] (a2.south) -- ++(0,-0.3) -| (ver.north);
    \draw[arw] (an.south) -- ++(0,-0.3) -| ([xshift=0.4cm]ver.north);

    \node[stagelbl] at (-1.5, -2.9) {\rotatebox{90}{Validation}};

  \end{tikzpicture}
  \caption{The MEP methodology. Three sequential preparation phases produce structured artifacts consumed by parallel agents. Task decomposition distributes independent work units to $n$ concurrent agents, whose outputs converge in integration verification.}
  \label{fig:methodology}
\end{figure}
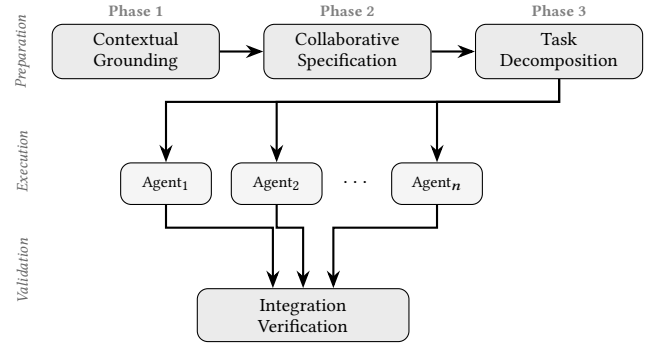

We propose mise en place (MEP) as a structured preparation
methodology for agentic coding. We observe in AI-assisted software development the same
dynamic that governs professional kitchens: practitioners who invest
in deliberate preparation---externalizing domain knowledge,
producing detailed specifications, and decomposing work into
structured task records---execute faster, with more alignment, and
require less corrective iteration. MEP formalizes preparation into
three sequential phases, each producing artifacts that feed the
next. Inspired by backward design~\cite{wiggins1998understanding},
the phases are completed before implementation begins.
Figure~\ref{fig:methodology} illustrates the phases and their
relationship to agent execution.

\subsection{Phase 1: Contextual Grounding}

The first phase externalizes domain expertise and tacit knowledge
into structured documents agents can consume---what
Polanyi~\cite{polanyi1966tacit} called tacit knowledge,
the understanding practitioners possess but struggle to articulate.
The artifacts are briefing documents (markdown files encoding domain
knowledge, competitive analysis, design philosophy) that force the
practitioner to articulate implicit knowledge and create a
persistent context layer agents reference throughout implementation.
Following backward design~\cite{wiggins1998understanding},
contextual grounding begins with outcomes rather than features.

\subsection{Phase 2: Collaborative Specification}

The second phase produces detailed design artifacts through
structured human-agent dialogue: the practitioner describes intent,
the agent proposes details, the practitioner accepts, rejects, or
modifies. The key mechanism is encoding value judgments as
specification constraints---what to \emph{exclude} is as important
as what to include. The output is a design document specifying
screens, interactions, data flows, and quality standards, capturing
not just the \emph{what} but the \emph{why}. This rationale enables
agents to make aligned micro-decisions during implementation
without escalating to the practitioner.

\subsection{Phase 3: Task Decomposition}

The final preparatory phase converts the specification into
structured, dependency-aware task records. We employ
Beads~\cite{yegge2025beads, emanuel2025beadsrust}---lightweight JSON
records backed by Git, carrying priorities, dependencies, and
acceptance criteria---though the principle generalizes. Fine-grained
decomposition enables parallel agent execution without coordination
overhead: when each task has clear boundaries and explicit
dependencies, multiple agents work simultaneously. The coordination
burden shifts from runtime to preparation time, where human
judgment about system architecture is most valuable. Following
parallel execution, integration verification validates outputs
against the specification.

The three phases draw individually on established
work~\cite{polanyi1958personal, horthy2025rpi, huntley2025ralph};
MEP's contribution is their integration into a phase-gated process
that completes before implementation begins.

%% file: sections/case-study.tex
\section{Case Study: Competitive Hackathon}
\label{sec:casestudy}

We report on the application of MEP during a competitive hackathon
to illustrate the methodology's behavior under realistic constraints. The
setting imposed fixed time limits, external evaluation, and a requirement to
produce working software---conditions that favor rapid iteration and
penalize unproductive preparation.

\subsection{Setting}

The hackathon took place in January 2026, organized around a publisher's
editorial archive and an AI-powered content API. Approximately 12 teams
competed in a five-hour window to build working prototypes, with judging
by an industry panel and a \$5{,}000 prize for the top project. Teams
self-organized around pitches submitted before the event~\cite{event2026hackathon}.
The practitioner-researcher had professional backgrounds in both
education and software engineering, a combination that informed the
product concept and design decisions throughout the event.

\subsection{Preparation Phase}

While most teams began coding within the first fifteen minutes, the
practitioner-researcher spent approximately two hours in deliberate
preparation, applying the three phases described in
Section~\ref{sec:methodology}.

\textbf{Contextual grounding} produced 10 planning documents totaling 9,386
words, including API exploration notes, competitive analysis, and---critically---an
extended dictation on pedagogical design philosophy drawn from the practitioner's
teaching experience, encoding tacit knowledge about how effective learning
environments create conditions for student thinking~\cite{polanyi1966tacit}.

\textbf{Collaborative specification} occurred concurrently, as the practitioner
refined the product concept through dialogue with an AI agent. The resulting
specification described a platform where teachers curate bounded research
environments from The Atlantic's archive and students conduct research using a
Socratic AI tutor. Key value judgments encoded in the specification included
constraining the product to a single assignment type (prioritizing depth over
breadth for a three-minute demo),
insisting on real API data rather than mocks, and scoping the analytics dashboard
to show the student's research \emph{journey} rather than only the final submission.

\textbf{Task decomposition} converted the specification into 64 structured
task records with priorities and dependencies. The planning-to-code
ratio---9,386 words of planning against 8,496 lines of source code---was
1.10:1. Table~\ref{tab:timeline} summarizes the timeline alongside the
deployment commit cadence.

\begin{table}[t]
  \centering
  \caption{Hackathon timeline, artifacts, and deployment cadence. Phases
    overlap inside the preparation block; agent implementation runs in
    parallel across four agents. The deployment-and-polish window
    produces nine commits at $\sim$5~minute mean intervals.}
  \label{tab:timeline}
  \footnotesize
  \begin{tabular}{@{}lrl@{}}
    \toprule
    \textbf{Phase} & \textbf{Duration} & \textbf{Key Artifacts / Cadence} \\
    \midrule
    Contextual grounding   & $\sim$2 hr      & 10 docs (9,386 words) \\
    Collab.\ specification & (concurrent)    & Product specification \\
    Task decomposition     & (concurrent)    & 64 beads w/ dependencies \\
    Agent implementation   & 184 min         & 43 beads closed (med.\ 5.9 min) \\
    \midrule
    Deploy: setup        & 22:24--22:39    & 5 commits, 1 every 3 min \\
    Deploy: gap          & 22:39--22:49    & --- \\
    Deploy: content fixes & 22:49--23:17    & 4 commits, 1 every 7 min \\
    \bottomrule
  \end{tabular}
\end{table}

\subsection{Implementation and Results}

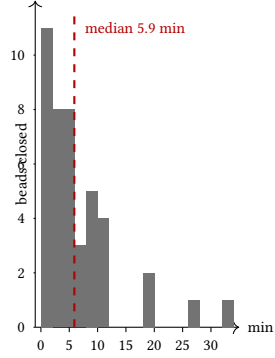
\begin{figure}[t]
  \centering
  \begin{tikzpicture}[
    yscale=0.36, xscale=0.075,
    every node/.style={font=\scriptsize},
  ]
    \draw[->] (-1,0) -- (35,0) node[right, font=\scriptsize] {min};
    \draw[->] (-1,0) -- (-1,12);
    \foreach \x in {0,5,10,15,20,25,30}
      \draw (\x,0) -- (\x,-0.18) node[below, font=\scriptsize] {\x};
    \foreach \y/\lab in {0/0,2/2,4/4,6/6,8/8,10/10}
      \draw (-1,\y) -- (-1.3,\y) node[left, font=\scriptsize] {\lab};
    \fill[black!55] (0,0)  rectangle (2,11);
    \fill[black!55] (2,0)  rectangle (4,8);
    \fill[black!55] (4,0)  rectangle (6,8);
    \fill[black!55] (6,0)  rectangle (8,3);
    \fill[black!55] (8,0)  rectangle (10,5);
    \fill[black!55] (10,0) rectangle (12,4);
    \fill[black!55] (18,0) rectangle (20,2);
    \fill[black!55] (26,0) rectangle (28,1);
    \fill[black!55] (32,0) rectangle (34,1);
    \draw[dashed, thick, red!70!black] (5.9,0) -- (5.9,11.6);
    \node[red!70!black, anchor=west, font=\scriptsize] at (6.3,11) {median 5.9 min};
    \node[rotate=90, font=\scriptsize] at (-3.5,6) {beads closed};
  \end{tikzpicture}
  \caption{Per-bead completion time distribution across the 43 closed
    hackathon beads. Most work units complete inside a single 10-minute
    agent session; long-tail outliers correspond to API-client and
    state-persistence tasks executed in parallel with shorter
    decomposed work.}
  \label{fig:beadtimes}
\end{figure}

Four parallel subagents were deployed across distinct feature areas:
classroom creation, student research workspace, day-one/\allowbreak
day-thirty demo toggle, and Socratic AI tutor. The agents drew on the
contextual grounding documents and specification produced during
preparation. The 64 task records served as the interface between
preparation and execution (Figure~\ref{fig:methodology}); each bead
encoded an independently executable unit with boundaries that required
no inter-agent coordination. Closure timestamps confirm genuine
parallelism: agents completed work simultaneously across feature
areas, with a median closure time of 5.9~minutes per bead
(Figure~\ref{fig:beadtimes}). The subsequent deployment phase produced
9~commits over 52~minutes. The final codebase comprised 43
TypeScript/TSX files totaling 8,496 lines, deployed to production as a
full-stack educational platform.

Three observations bear on evaluation. First, the architecture required no
structural refactoring during deployment: the bugs that emerged were
integration and styling issues (API content truncation, inter-component
data flow, favicon configuration), not architectural misalignment.
Second, the preparation-to-execution ratio was approximately 5.7:1
(two hours of preparation against $\sim$21 minutes of active
implementation per agent across four parallel agents). Bug-type beads
resolved at a median 1.2~minutes (mean 1.5~min) versus 9.7~minutes for
implementation tasks, suggesting parallel-execution defects were
detected and corrected quickly.
Third, near-zero architectural rework is consistent with the hypothesis
that rich upfront context reduces agent misalignment, though we cannot
establish causation from a single case study.

\subsection{Variation Across Teams}
\label{sec:teamvariation}

The 12-team field offers a qualitative sketch of how peers approached
the same constraints. We did not instrument other teams; identities are
anonymized and observations are coarse-grained, drawing on publicly
shared pitches and event observation. Table~\ref{tab:teams} groups
pitches into thematic clusters with each team's described workflow style.

\begin{table}[t]
  \centering
  \caption{Twelve-team field at the hackathon, grouped by problem
    cluster, with workflow style as inferred from pitches and
    observation. Identifiers are anonymized; ``vibe'' denotes a
    self-described non-technical or iterative-prompting workflow,
    ``decomp.'' denotes explicit planning before coding, and ``mixed''
    denotes both modes within one team.}
  \label{tab:teams}
  \scriptsize
  \begin{tabular}{@{}p{0.06\columnwidth}p{0.46\columnwidth}p{0.16\columnwidth}p{0.12\columnwidth}@{}}
    \toprule
    \textbf{ID} & \textbf{Pitch theme} & \textbf{Cluster} & \textbf{Workflow} \\
    \midrule
    T1  & Infinite-canvas archive graph        & Discovery       & decomp.\ \\
    T2  & Reading-comprehension overlay        & Education       & vibe \\
    T3  & In-article excerpt exploration       & Discovery       & mixed \\
    T4  & AI ``knowledge gauge'' (reading level) & Accessibility & vibe \\
    T5  & Auto-generated event timelines       & Context         & vibe \\
    T6  & Topic-following / catch-up alerts    & Personalization & mixed \\
    T7  & Editorial content optimization (CORE) & Personalization & decomp.\ \\
    T8  & Reader-interpretation probes         & Trust           & decomp.\ \\
    T9  & Structured reader-expert input       & Trust           & decomp.\ \\
    T10 & Argument-evolution interaction layer & Discovery       & decomp.\ \\
    T11 & Newsroom knowledge-graph copilot     & Productivity    & decomp.\ \\
    T12 & Curated classroom research workspace (this study) & Education & decomp.\ \\
    \bottomrule
  \end{tabular}
\end{table}

Three patterns emerge. \textbf{Workflow style} split roughly in half:
vibe-coding pitches (T2, T4, T5) leaned on iterative prompting, while
engineering-led teams (T1, T7--T12) opened with explicit planning
artifacts. \textbf{Single- versus multi-agent} use trended toward
single-agent loops; only T12 (this study) fanned parallel subagents
from a decomposed plan. \textbf{Artifact production} was correspondingly
uneven across the two clusters. Without controlled comparison we
cannot link these patterns to outcomes, but the field samples the
preparation-versus-iteration spectrum motivating RQ1 and the
multi-team replication study in Section~\ref{sec:agenda}.

%% file: sections/context-fluency.tex
\section{Context Fluency: An Emerging Skill}

Effective agentic coding requires a skill set distinct from both
traditional programming and prompt engineering. We propose
\emph{context fluency} to describe this capability: the ability to
create rich, structured context that AI agents can act on. Context
fluency captures domain expertise, value judgments, and design intent
in machine-legible form---not as one-shot prompts, but as persistent
informational environments. Prompt engineering optimizes individual
instructions; context fluency is \emph{upstream} of prompting,
concerned with the informational architecture surrounding agent
execution~\cite{anthropic2025context, gartner2025context}. Where MEP
describes a sequence of phases producing artifacts, context fluency
describes the practitioner skill that makes the process effective.

We identify four components of context fluency:
\begin{description}
  \item[Decomposition:] Breaking problems into discrete, parallelizable
    tasks with clear boundaries, enabling concurrent agent execution.
  \item[Specification:] Describing not just \emph{what} to build but
    \emph{why}, so agents can make aligned micro-decisions without human
    intervention.
  \item[Constraint definition:] Knowing what to exclude, simplify, or
    defer---scope management as a first-class concern.
  \item[Domain encoding:] Externalizing tacit
    knowledge~\cite{polanyi1966tacit} that agents cannot generate on
    their own---directly instantiating what Nonaka and
    Takeuchi~\cite{nonaka1995knowledge} term \emph{externalization},
    converting tacit knowledge into explicit, communicable form.
    Twenty minutes of dictated pedagogical intuitions
    produced context that substantially reduced iteration on the
    domain-specific tutor component.
\end{description}

These components map to established frameworks. Backward
design~\cite{wiggins1998understanding} contributes outcome-driven
preparation; Polanyi's tacit knowledge~\cite{polanyi1958personal}
frames the externalization challenge---``we know more than we can
tell,'' and context fluency is the discipline of telling it anyway.
Context fluency also resembles pedagogical scaffolding: structuring
an environment so the learner (or agent) can act independently.

If context fluency is a distinct skill, the implications extend
beyond individual practice. For developer education, curricula should
cultivate specification, decomposition, and domain-encoding alongside
programming competencies. For hiring, practitioners with strong
domain knowledge and pedagogical instincts may be disproportionately
effective in agentic workflows~\cite{dakhel2025humanai}. For tool
design, preparation aids---task systems~\cite{yegge2025beads,
emanuel2025beadsrust}, specification frameworks~\cite{github2025speckit},
and context engineering platforms~\cite{anthropic2025context}---are
essential infrastructure, not optional enhancements.

%% file: sections/research-agenda.tex
\section{Limitations, Research Agenda, and Conclusion}
\label{sec:agenda}

\subsection{Limitations}
\label{sec:limitations}

The empirical claims rest on a single hackathon, a single
practitioner, and no control group. (i)~\textbf{Single hackathon,
single case study.} Findings are illustrative, not generalizable; a
five-hour competitive setting is not multi-month software
development. (ii)~\textbf{No control group, no iteration-first
comparison.} We did not run a paired condition without preparation,
nor instrument the other 11 teams; the 5.7:1 ratio and near-zero
rework are observations, not effects against a baseline.
(iii)~\textbf{Operator expertise as a confound.} The practitioner's
dual background in education and software engineering likely
contributed to the contextual grounding's quality, so we cannot
separate the methodology's contribution from operator expertise. The
team-variation sketch in Section~\ref{sec:teamvariation} adds context
but does not constitute controlled comparison.

\subsection{Research Agenda}

We propose five research questions and three concrete research
lines that follow from the limitations above:

\begin{description}
  \item[RQ1:] \emph{Does deliberate preparation reduce agent
    misalignment compared to iterative development?} A
    \textbf{comparative study} pairing preparation-first and
    iteration-first workflows on matched tasks would establish
    whether the upfront investment yields net savings (addresses
    Limitation~(ii)).

  \item[RQ2:] \emph{What constitutes ``sufficient'' context---can we
    define stopping criteria for the preparation phase?} A
    saturation-style analysis of alignment quality vs.\ context
    volume would help practitioners know when to stop preparing.

  \item[RQ3:] \emph{How does context fluency vary across
    practitioners, domains, and agent models?} A \textbf{multi-team
    replication study} generalizing the sketch in
    Section~\ref{sec:teamvariation} could reveal whether context
    fluency is trainable or a function of prior domain
    knowledge~\cite{polanyi1958personal} (addresses
    Limitation~(iii)).

  \item[RQ4:] \emph{Can MEP scale beyond prototyping to multi-month
    development?} A \textbf{longitudinal study} of context-fluency
    development would test scale
    applicability~\cite{dakhel2025humanai} (addresses
    Limitation~(i)).

  \item[RQ5:] \emph{What is the relationship between preparation
    time and implementation quality?} A multi-project dataset could
    yield quantitative models of the preparation-quality
    relationship~\cite{alomar2025promptware}.
\end{description}

The three named lines---comparative, replication, and
longitudinal---convert the empirical concerns into a falsifiable
program; RQ1 is the most immediately tractable.

\subsection{Artifact Availability}
\label{sec:artifact}

To support the SIGSOFT Open Science Policy and enable replication, we
release a research artifact bundle containing the anonymized hackathon
data (timing, commit, and bead records reported in this paper), the
prompts and agent configurations used, and the context-engineering
scaffolds (CLAUDE.md, beads schema, planning templates) that
constitute the MEP harness. The bundle is licensed CC-BY~4.0 (data) and
MIT (code). It is archived on Zenodo at
\url{https://doi.org/10.5281/zenodo.19868258}.

\subsection{Conclusion}

MEP is a starting framework---a formalization of practices that
experienced practitioners converge on
independently~\cite{huntley2025ralph, horthy2025rpi, yegge2025beads},
grounded in what Nonaka and Takeuchi~\cite{nonaka1995knowledge} term
externalization. We claim no generalizability from a single
hackathon, but argue that the alignment problem in agentic coding is
fundamentally a preparation problem, deserving treatment as a
first-class engineering discipline. Context fluency names the
emerging practitioner skill at the intersection of domain expertise
and specification craft~\cite{anthropic2025context,
github2025speckit}.

%% file: main.bbl

\begin{thebibliography}{22}


\ifx \showCODEN    \undefined \def \showCODEN     #1{\unskip}     \fi
\ifx \showISBNx    \undefined \def \showISBNx     #1{\unskip}     \fi
\ifx \showISBNxiii \undefined \def \showISBNxiii  #1{\unskip}     \fi
\ifx \showISSN     \undefined \def \showISSN      #1{\unskip}     \fi
\ifx \showLCCN     \undefined \def \showLCCN      #1{\unskip}     \fi
\ifx \shownote     \undefined \def \shownote      #1{#1}          \fi
\ifx \showarticletitle \undefined \def \showarticletitle #1{#1}   \fi
\ifx \showURL      \undefined \def \showURL       {\relax}        \fi
\providecommand\bibfield[2]{#2}
\providecommand\bibinfo[2]{#2}
\providecommand\natexlab[1]{#1}
\providecommand\showeprint[2][]{arXiv:#2}

\bibitem[Alomar et~al\mbox{.}(2025)]%
        {alomar2025promptware}
\bibfield{author}{\bibinfo{person}{Eman~Abdullah Alomar} {et~al\mbox{.}}}
  \bibinfo{year}{2025}\natexlab{}.
\newblock \showarticletitle{Promptware Engineering: The New Frontier of
  Software Engineering}.
\newblock \bibinfo{journal}{\emph{arXiv preprint}}
  \bibinfo{volume}{arXiv:2503.02400} (\bibinfo{year}{2025}).
\newblock
\href{https://doi.org/10.48550/arXiv.2503.02400}{doi:\nolinkurl{10.48550/arXiv.2503.02400}}


\bibitem[{Anthropic}(2025)]%
        {anthropic2025context}
\bibfield{author}{\bibinfo{person}{{Anthropic}}.}
  \bibinfo{year}{2025}\natexlab{}.
\newblock \bibinfo{title}{Effective Context Engineering for AI Agents}.
\newblock \bibinfo{howpublished}{Anthropic Documentation}.
\newblock
\urldef\tempurl%
\url{https://docs.anthropic.com/en/docs/build-with-claude/context-engineering}
\showURL{%
\tempurl}


\bibitem[Brown et~al\mbox{.}(2020)]%
        {brown2020gpt3}
\bibfield{author}{\bibinfo{person}{Tom~B. Brown}, \bibinfo{person}{Benjamin
  Mann}, \bibinfo{person}{Nick Ryder}, \bibinfo{person}{Melanie Subbiah},
  \bibinfo{person}{Jared Kaplan}, \bibinfo{person}{Prafulla Dhariwal},
  \bibinfo{person}{Arvind Neelakantan}, \bibinfo{person}{Pranav Shyam},
  \bibinfo{person}{Girish Sastry}, \bibinfo{person}{Amanda Askell},
  {et~al\mbox{.}}} \bibinfo{year}{2020}\natexlab{}.
\newblock \showarticletitle{Language Models are Few-Shot Learners}. In
  \bibinfo{booktitle}{\emph{Advances in Neural Information Processing Systems
  (NeurIPS)}}.
\newblock


\bibitem[Dakhel et~al\mbox{.}(2025)]%
        {dakhel2025humanai}
\bibfield{author}{\bibinfo{person}{Arghavan~Moradi Dakhel} {et~al\mbox{.}}}
  \bibinfo{year}{2025}\natexlab{}.
\newblock \showarticletitle{A Taxonomy of Human-AI Collaboration in Software
  Engineering}.
\newblock \bibinfo{journal}{\emph{arXiv preprint}}
  \bibinfo{volume}{arXiv:2501.08774} (\bibinfo{year}{2025}).
\newblock
\href{https://doi.org/10.48550/arXiv.2501.08774}{doi:\nolinkurl{10.48550/arXiv.2501.08774}}


\bibitem[Emanuel(2025)]%
        {emanuel2025beadsrust}
\bibfield{author}{\bibinfo{person}{Jeffrey Emanuel}.}
  \bibinfo{year}{2025}\natexlab{}.
\newblock \bibinfo{title}{beads\_rust: High-Performance Task Tracking for AI
  Agents}.
\newblock \bibinfo{howpublished}{GitHub repository}.
\newblock
\urldef\tempurl%
\url{https://github.com/Dicklesworthstone/beads_rust}
\showURL{%
\tempurl}


\bibitem[{Gartner}(2025)]%
        {gartner2025context}
\bibfield{author}{\bibinfo{person}{{Gartner}}.}
  \bibinfo{year}{2025}\natexlab{}.
\newblock \bibinfo{title}{Context Engineering Is In, Prompt Engineering Is
  Out}.
\newblock \bibinfo{howpublished}{Gartner Research Note}.
\newblock
\newblock
\shownote{Gartner identified context engineering as a key shift in AI
  application development for 2025}.


\bibitem[{GitHub}(2025)]%
        {github2025speckit}
\bibfield{author}{\bibinfo{person}{{GitHub}}.} \bibinfo{year}{2025}\natexlab{}.
\newblock \bibinfo{title}{Spec Kit: Specification-Driven Development for AI
  Agents}.
\newblock \bibinfo{howpublished}{GitHub repository}.
\newblock
\urldef\tempurl%
\url{https://github.com/github/spec-kit}
\showURL{%
\tempurl}


\bibitem[{Hacks/Hackers and The Atlantic and Infactory}(2026)]%
        {event2026hackathon}
\bibfield{author}{\bibinfo{person}{{Hacks/Hackers and The Atlantic and
  Infactory}}.} \bibinfo{year}{2026}\natexlab{}.
\newblock \bibinfo{title}{Building Future {AI} News Experiences with The
  Atlantic and Infactory}.
\newblock \bibinfo{howpublished}{Hackathon event description}.
\newblock
\newblock
\shownote{Approximately 12 teams, five-hour build window, \$5{,}000 prize for
  top project}.


\bibitem[Horthy(2025)]%
        {horthy2025rpi}
\bibfield{author}{\bibinfo{person}{Dex Horthy}.}
  \bibinfo{year}{2025}\natexlab{}.
\newblock \bibinfo{title}{The {RPI} Methodology: Research, Plan, Implement for
  Agentic Coding}.
\newblock \bibinfo{howpublished}{HumanLayer Blog}.
\newblock
\urldef\tempurl%
\url{https://humanlayer.dev/blog/rpi-methodology}
\showURL{%
\tempurl}


\bibitem[Huntley(2025)]%
        {huntley2025ralph}
\bibfield{author}{\bibinfo{person}{Geoffrey Huntley}.}
  \bibinfo{year}{2025}\natexlab{}.
\newblock \bibinfo{title}{The Ralph Loop: A Pattern for AI Agent Iteration}.
\newblock \bibinfo{howpublished}{Blog post}.
\newblock
\urldef\tempurl%
\url{https://ghuntley.com/ralph/}
\showURL{%
\tempurl}


\bibitem[Karpathy(2025)]%
        {karpathy2025vibe}
\bibfield{author}{\bibinfo{person}{Andrej Karpathy}.}
  \bibinfo{year}{2025}\natexlab{}.
\newblock \bibinfo{title}{Vibe Coding}.
\newblock \bibinfo{howpublished}{Twitter/X post}.
\newblock
\urldef\tempurl%
\url{https://x.com/karpathy/status/1886192184808149383}
\showURL{%
\tempurl}
\newblock
\shownote{Term entered Merriam-Webster and Collins Dictionary in 2025}.


\bibitem[Liu et~al\mbox{.}(2023)]%
        {liu2023prompt}
\bibfield{author}{\bibinfo{person}{Pengfei Liu}, \bibinfo{person}{Weizhe Yuan},
  \bibinfo{person}{Jinlan Fu}, \bibinfo{person}{Zhengbao Jiang},
  \bibinfo{person}{Hiroaki Hayashi}, {and} \bibinfo{person}{Graham Neubig}.}
  \bibinfo{year}{2023}\natexlab{}.
\newblock \showarticletitle{Pre-train, Prompt, and Predict: A Systematic Survey
  of Prompting Methods in Natural Language Processing}.
\newblock \bibinfo{journal}{\emph{Comput. Surveys}} \bibinfo{volume}{55},
  \bibinfo{number}{9} (\bibinfo{year}{2023}), \bibinfo{pages}{1--35}.
\newblock
\href{https://doi.org/10.1145/3560815}{doi:\nolinkurl{10.1145/3560815}}


\bibitem[Mozannar et~al\mbox{.}(2023)]%
        {developer2023mental}
\bibfield{author}{\bibinfo{person}{Hussein Mozannar} {et~al\mbox{.}}}
  \bibinfo{year}{2023}\natexlab{}.
\newblock \showarticletitle{Reading Between the Lines: Modeling User Behavior
  and Costs in AI-Assisted Programming}.
\newblock \bibinfo{journal}{\emph{arXiv preprint}}
  \bibinfo{volume}{arXiv:2210.14306} (\bibinfo{year}{2023}).
\newblock
\href{https://doi.org/10.48550/arXiv.2210.14306}{doi:\nolinkurl{10.48550/arXiv.2210.14306}}


\bibitem[Nonaka and Takeuchi(1995)]%
        {nonaka1995knowledge}
\bibfield{author}{\bibinfo{person}{Ikujiro Nonaka} {and}
  \bibinfo{person}{Hirotaka Takeuchi}.} \bibinfo{year}{1995}\natexlab{}.
\newblock \bibinfo{booktitle}{\emph{The Knowledge-Creating Company: How
  Japanese Companies Create the Dynamics of Innovation}}.
\newblock \bibinfo{publisher}{Oxford University Press}, \bibinfo{address}{New
  York}.
\newblock


\bibitem[Peng et~al\mbox{.}(2023)]%
        {peng2023copilot}
\bibfield{author}{\bibinfo{person}{Sida Peng}, \bibinfo{person}{Eirini
  Kalliamvakou}, \bibinfo{person}{Peter Cihon}, {and} \bibinfo{person}{Mert
  Demirer}.} \bibinfo{year}{2023}\natexlab{}.
\newblock \showarticletitle{The Impact of AI on Developer Productivity:
  Evidence from GitHub Copilot}.
\newblock \bibinfo{journal}{\emph{arXiv preprint}}
  \bibinfo{volume}{arXiv:2302.06590} (\bibinfo{year}{2023}).
\newblock
\href{https://doi.org/10.48550/arXiv.2302.06590}{doi:\nolinkurl{10.48550/arXiv.2302.06590}}


\bibitem[Polanyi(1958)]%
        {polanyi1958personal}
\bibfield{author}{\bibinfo{person}{Michael Polanyi}.}
  \bibinfo{year}{1958}\natexlab{}.
\newblock \bibinfo{booktitle}{\emph{Personal Knowledge: Towards a Post-Critical
  Philosophy}}.
\newblock \bibinfo{publisher}{University of Chicago Press},
  \bibinfo{address}{Chicago, IL}.
\newblock


\bibitem[Polanyi(1966)]%
        {polanyi1966tacit}
\bibfield{author}{\bibinfo{person}{Michael Polanyi}.}
  \bibinfo{year}{1966}\natexlab{}.
\newblock \bibinfo{booktitle}{\emph{The Tacit Dimension}}.
\newblock \bibinfo{publisher}{University of Chicago Press},
  \bibinfo{address}{Chicago, IL}.
\newblock


\bibitem[Vasconcelos et~al\mbox{.}(2024)]%
        {facct2024trust}
\bibfield{author}{\bibinfo{person}{Marisa Vasconcelos}, \bibinfo{person}{Jack
  Jamieson}, \bibinfo{person}{Umang Bhatt}, {and} \bibinfo{person}{Q.~Vera
  Liao}.} \bibinfo{year}{2024}\natexlab{}.
\newblock \showarticletitle{Understanding Trust in AI-Assisted Code
  Generation}. In \bibinfo{booktitle}{\emph{Proceedings of the ACM Conference
  on Fairness, Accountability, and Transparency (FAccT)}}.
  \bibinfo{publisher}{ACM}, \bibinfo{address}{New York, NY}.
\newblock


\bibitem[{Veracode}(2025)]%
        {veracode2025ai}
\bibfield{author}{\bibinfo{person}{{Veracode}}.}
  \bibinfo{year}{2025}\natexlab{}.
\newblock \bibinfo{title}{State of Software Security: The Rise of AI Code}.
\newblock \bibinfo{howpublished}{Industry report}.
\newblock
\newblock
\shownote{Reports that 45\% of AI-generated code contains security flaws, with
  a 10x spike in security findings from AI-generated code by June 2025}.


\bibitem[Wei et~al\mbox{.}(2022)]%
        {wei2022chain}
\bibfield{author}{\bibinfo{person}{Jason Wei}, \bibinfo{person}{Xuezhi Wang},
  \bibinfo{person}{Dale Schuurmans}, \bibinfo{person}{Maarten Bosma},
  \bibinfo{person}{Brian Ichter}, \bibinfo{person}{Fei Xia},
  \bibinfo{person}{Ed~H. Chi}, \bibinfo{person}{Quoc~V. Le}, {and}
  \bibinfo{person}{Denny Zhou}.} \bibinfo{year}{2022}\natexlab{}.
\newblock \showarticletitle{Chain-of-Thought Prompting Elicits Reasoning in
  Large Language Models}. In \bibinfo{booktitle}{\emph{Advances in Neural
  Information Processing Systems (NeurIPS)}}.
\newblock


\bibitem[Wiggins and McTighe(1998)]%
        {wiggins1998understanding}
\bibfield{author}{\bibinfo{person}{Grant Wiggins} {and} \bibinfo{person}{Jay
  McTighe}.} \bibinfo{year}{1998}\natexlab{}.
\newblock \bibinfo{booktitle}{\emph{Understanding by Design}}.
\newblock \bibinfo{publisher}{Association for Supervision and Curriculum
  Development}, \bibinfo{address}{Alexandria, VA}.
\newblock


\bibitem[Yegge(2025)]%
        {yegge2025beads}
\bibfield{author}{\bibinfo{person}{Steve Yegge}.}
  \bibinfo{year}{2025}\natexlab{}.
\newblock \bibinfo{title}{Beads: External Memory for AI Agents}.
\newblock \bibinfo{howpublished}{GitHub repository}.
\newblock
\urldef\tempurl%
\url{https://github.com/steveyegge/beads}
\showURL{%
\tempurl}


\end{thebibliography}
